\documentclass[numberedappendix,apj,twocolumn]{emulateapj}

\newcommand{\lya}{Ly$\alpha$}

\newcommand{\zdrop}{$z_{850}$-dropout }

\newcommand{\zFilter}{$z_{850}$ }
\newcommand{\jFilter}{$J_{110}$ }
\newcommand{\hFilter}{$H_{160}$ }

\newcommand{\iF}{$i_{775}$}
\newcommand{\zF}{$z_{850}$}
\newcommand{\vF}{$V_{606}$}
\newcommand{\jF}{$J_{110}$}
\newcommand{\hF}{$H_{160}$}
\newcommand{\like}{{\cal L}}

\shorttitle{UDF05: $z_{850}$-dropouts}
\shortauthors{Oesch et al.}

\begin{document}

\title{The UDF05 Follow-up of the HUDF: II. Constraints on Reionization from $z$-dropout Galaxies
\altaffilmark{1}}

\altaffiltext{1}{Mostly based on data obtained with the \textit{Hubble Space Telescope} operated by AURA, Inc. for NASA under contract NAS5-26555. Partly based on data from the \textit{Spitzer Space Telescope} operated by Jet Propulsion Laboratory, California Institute of Technology under NASA contract 1407.}

\author{P. A. Oesch\altaffilmark{2}, 
C. M. Carollo\altaffilmark{2}, 
M. Stiavelli\altaffilmark{3}, 
M. Trenti\altaffilmark{3}, 
L. E. Bergeron\altaffilmark{3}, 
A. M. Koekemoer\altaffilmark{3},
R. A. Lucas\altaffilmark{3}, 
C. M. Pavlovsky\altaffilmark{3}, 
S. V. W. Beckwith\altaffilmark{3},
T. Dahlen\altaffilmark{3},
H. C. Ferguson\altaffilmark{3},
Jonathan P. Gardner\altaffilmark{4},
S. J. Lilly\altaffilmark{2},  
B. Mobasher\altaffilmark{3},
N. Panagia\altaffilmark{3,5,6}
}

\altaffiltext{2}{Department of Physics, Institute of Astronomy, Eidgen\"{o}ssische Technische Hochschule (ETH Zurich), CH-8093 Zurich, Switzerland; poesch@phys.ethz.ch}
\altaffiltext{3}{Space Telescope Science Institute, Baltimore, MD 21218, United States}
\altaffiltext{4}{Laboratory for Observational Cosmology, Code 665, NASA's Goddard Space
Flight Center, Greenbelt MD 20771}
\altaffiltext{5}{INAF- Osservatorio Astrofisico di Catania, Via S. Sofia 78, I-95123 Catania, Italy}
\altaffiltext{6}{Supernova Ltd., OYV 131, Northsound Road, Virgin Gorda, British Virgin Islands}

\begin{abstract}
We detect three (plus one less certain) $z_{850}$-dropout sources in two separate fields (HUDF and NICP34) of our UDF05 HST NICMOS images. These $z\sim7$ Lyman-Break Galaxy (LBG) candidates allow us to constrain the Luminosity Function (LF) of the star forming galaxy population at those epochs. By assuming a change in only $M_*$ and adopting a linear evolution in redshift, anchored to the measured values at $z\sim6$, the best-fit evolution coefficient is found to be $0.43\pm0.19$ mag per unit redshift ($0.36\pm0.18$, if including all four candidates), which provides a value of $M_*(z=7.2)=-19.7\pm0.3$. This implies a drop of the luminosity density in LBGs by a factor of $\sim2-2.5$ over the $\sim$ 170 Myr that separate  $z\sim6$ and $z\sim7$, and a steady evolution for the LBG LF out to $z\sim7$, at the same rate that is observed throughout the $z\sim3$ to 6 period. This puts a strong constraint on the star-formation histories of $z\sim6$ galaxies, whose ensemble star-formation rate density must be lower by a factor 2 at $\sim$170 Myr before the epoch at which they are observed. In particular, a large fraction of stars in the $z\sim6$ LBG population must form at redshifts well above $z\sim7$. The rate of ionizing photons produced by the LBG population decreases consistently with the decrease in the cosmic star formation rate density. Extrapolating this steady evolution of the LF out to higher redshifts, we estimate that galaxies would be able to reionize the universe by $z\sim6$, provided that the faint-end slope of the $z>7$ LF steepens to $\alpha\sim-1.9$, and that faint galaxies, with luminosities below the current detection limits, contribute a substantial fraction of the required ionizing photons. This scenario gives however an integrated optical depth to electron scattering that is $\sim2 \sigma$ below the WMAP-5 measurement. Therefore, altogether, our results indicate that, should galaxies be the primary contributors to reionization, either the currently detected evolution of the galaxy population slows down at $z\gtrsim7$, or the LF evolution must be compensated by a decrease in metallicity and a corresponding increase in ionization efficiency at these early epochs. 

\end{abstract}


\section{Introduction}
One of the most fundamental open issues in observational cosmology is what sources reionized the early Universe.  Since the $z\gtrsim6$ quasar Luminosity Function (LF) is found to be too shallow to provide enough ionizing photons \citep[e.g.][]{fan01,meik05,shan07}, the most likely culprits are thought to be galaxies.  A key question is therefore, how the star forming galaxy population evolves beyond the putative end of reionization at $z\sim6$, i.e., well into the epoch of reionization.

Direct searches for  Ly$\alpha$ emission  provide a powerful way to study high-redshift galaxies  \citep[e.g.][]{iye06,star07,ota08}; however, it is difficult to estimate the (continuum and) ionizing flux of Ly$\alpha$ emitters, and thus their contribution to  reionization.  A more fruitful approach is provided by the Lyman break technique.  

Thousands of Lyman-break galaxies (LBGs) have been identified between $z \sim$ 3 and 6 from optical images, providing an estimate for the rest-frame UV LF at those redshifts \citep[e.g. Oesch et al 2007, hereafter O07; Bouwens et al 2007, hereafter B07; see also][]{stei99,sawi06,yosh06,iwat07}.  The evolution of the LBG LF throughout this redshift range is however still being debated:  especially at the bright end, the strong clustering of bright galaxies and the tiny area coverage of space-based images prevent setting strong constraints on the LF. Another important factor is that the number of spectroscopically confirmed LBGs is still small due to the faintness of the sources. Within these caveats, O07 and B07 have published independent, self-consistent analyses, which agree on a remarkably steady evolution of $\Delta M_*\sim0.35$ mag per unit redshift over the $z=3-6$ epoch. 

At redshifts $z>6$, the rest-frame \lya\ line shifts into the near-infrared (NIR) and the relatively low efficiency and  small field of view of current NIR detectors have limited the identification of LBGs. Thus, in contrast to the large number of sources detected out to $z\sim6$, the numerous searches for $z\gtrsim7$ LBGs have led to only a handful of candidates \citep[e.g. Bouwens et al 2008, hereafter B08;][]{mann06,bouw06,Windhorst04,stan08}.  Searches for intrinsically faint $z\gtrsim7$ LBGs  around lensing clusters also resulted only in small numbers of candidates \citep[e.g.][]{rich06,brad08,rich08}, due to the small areas subject to the lensing magnification. 

One of the most comprehensive studies is the recent analysis of B08, who have  measured the $z\sim7$ LBG LF from a large compilation of both HST and ground-based data, covering a total area of $\sim$ 271 arcmin$^2$ over the two fields of the GOODS survey \citep{giav04}.  Again within the caveats imposed by probing, with small number statistics, only the bright-end of the LF, these authors report that the dimming of the LBG LF which has been measured out  to $z \sim 6$ continues  with no substantial change out to $z\sim7$. 

A summary of the previous work in the context of reionization is that, albeit within some restrictive assumptions, the detected  $z\sim6$ LBG population appears to be just barely able to maintain the Universe ionized \citep[e.g.][]{ferg02,lehn03,stia04,bunk04,bouw06,gned08}, and also the detected $z\sim7$ LBGs fall short of being able to provide the photons required to reionize the Universe \citep{bolt07}. It seems that fainter galaxies must have played a very important role \citep[e.g.][]{Windhorst04,rich08}. 

Here we present our independent derivation of constraints on the $z\sim7$ LBG LF. 
This is desirable in order to independently check the different effects of various uncertainties in the process of estimating LFs, including the data reduction, object detection as well as estimation of effective selection volumina.
We use our UDF05 data as well as the optical (Beckwith et al. 2007) and NIR (Thompson et al. 2005) data of the Hubble Ultra Deep Field (HUDF). The UDF05 survey is a 204-orbit HST Large Program of ultradeep ACS and NICMOS observations of multiple fields, each located $\sim$10$'$ away from HUDF (see O07). The UDF05  was designed to image with ACS-WFC the two NICMOS parallel fields (hereafter NICP12 and NICP34) that were acquired while the HUDF was observed with the ACS and thus facilitating a reliable search for $z>6$ galaxies in these fields to unprecedented depth. 

The combined NIR NICMOS dataset that we study in this paper consists of three fields, covering a total area of 7.9 arcmin$^2$, with 5$\sigma$ point-source magnitude limits varying between $27.9-28.8$ in F160W ($H_{160}$) and $27.6-28.6$ in F110W ($J_{110}$). The 50\% completeness limits vary between 26.9 and 28.1 mag (see Tab. \ref{tab:obslog} and \S \ref{sec:sim}). Each of the fields consists of one deep pointing and a slightly shallower flanking area (by $\sim 0.6$ mag). These data, also used in B08, provide the deepest NIR data available to date.

\begin{deluxetable}{lccccc}
\tablecaption{Depth and Area of the UDF05 and HUDF NICMOS Observations \label{tab:obslog}}
\tablewidth{0 pt}
\tablecolumns{5}
\small
\tablehead{\colhead{Field} & \colhead{$5 \sigma$\tablenotemark{a}} &\colhead{$5 \sigma$} & \colhead{$5 \sigma$} & \colhead{$C_{50\%}$\tablenotemark{b}} &\colhead{Area} \\
& \colhead{$z_{850}$} & \colhead{$J_{110}$} & \colhead{$H_{160}$} &\colhead{$H_{160}$} & \colhead{arcmin$^2$}}

\startdata
NICP34 deep &  28.3 &  28.6 &    28.8 & 28.1 &   0.7     \\
NICP34 shallow  & 28.2 & 27.9 &    28.1  & 27.3 &  0.4     \\

NICP12 deep  & 28.4  & 28.4 &    28.6 & 27.8 &  0.7     \\
NICP12 shallow & 28.4 & 27.8 &    28.0 & 27.3 &  0.4     \\

HUDF deep    &  28.6 & 28.2 &    28.2 & 27.5 &  0.7     \\ 
HUDF shallow  & 28.6 &  27.6 &  27.6 & 26.9 &  5.0

\enddata

\tablenotetext{a}{within an aperture of 0\farcs3 radius}
\tablenotetext{b}{50\% completeness limit, as estimated in section \S \ref{sec:sim}}

\end{deluxetable}

In \S \ref{sec:data}  we describe the data and the steps adopted to identify the $z \sim 7$ LBG candidates and constrain the LBG LF. In \S \ref{sec:discussion} we present our results and their implication for the evolution of the LBG LF into the reionization epoch, and in \S \ref{sec:discussion2} we discuss the contribution of the detected $z\sim7$ galaxies to reionization. We adopt the concordance cosmology defined by $\Omega_M=0.3, \Omega_\Lambda=0.7, H_0=70$ kms$^{-1}$Mpc$^{-1}$, i.e. $h=0.7$. Where necessary we assume $\Omega_b h^2 = 0.02265$ \citep{hins08}. Magnitudes are given in the AB system \citep{okeg83}.

\section{Observations and source selection}
\label{sec:data}

\subsection{The UDF05 and HUDF data}

We reprocessed the HUDF data  using  improved data reduction algorithms relative to the publicly released data, including an iterative sky subtraction scheme. 
All NICMOS data were  drizzled to a 0\farcs09 pixel scale. The optical ACS data were rebinned to this same scale and convolved with a Gaussian filter to match the width of the NICMOS point spread function (PSF). We used the \hF-band data to detect galaxies with the  SExtractor program \citep{bert96},  and measured their colors in elliptical apertures scaled to $1.5\times$ their Kron radii, matched to the detection image. Total magnitudes were measured in $2.5$ Kron (AUTO) apertures and were corrected for flux losses in the wings of the PSF ($0.07-0.10$ mag).

Only detections with signal-to-noise ratio (S/N) larger than 4.5 within 0\farcs3 radius apertures in the \hF-band were considered. Flux variations in apertures of different sizes, laid down on empty (sky) areas, were estimated in order to correct for  pixel-to-pixel noise correlation in the images; the rms maps were then rescaled so as to match the noise in 0\farcs3 radius apertures. Undetected fluxes were replaced with their 1$-\sigma$ upper limits.  The zeropoints of the \jFilter and \hFilter filters were corrected by 0.16 and 0.04 mag, respectively, in order to account for non-linearity of the NICMOS NIC3 detector (see de Jong 2006). Note that any additional effect of detector non-linearity would be minimal as our sample covers only a small range in magnitudes.

\subsection{The \zdrop candidates}

Candidate galaxies at redshifts between 6.8 and 8 were selected from the SExtractor catalog according to the following color criteria:
\begin{eqnarray*}
	(z_{850}-J_{110})&>&1.3 \\ 
	(z_{850}-J_{110})&>&1.3 + 0.4 (J_{110}-H_{160})  \\
	(J_{110}-H_{160})&<&1.2  \\
	S/N(H_{160}) > 4.5 \quad & \wedge & \quad S/N(J_{110})>2 \\
	S/N(V_{606}) <2 \quad & \wedge & \quad S/N(i_{775}) <2	
\end{eqnarray*}

\begin{figure}[t]
	\centering
		\includegraphics[width=\linewidth]{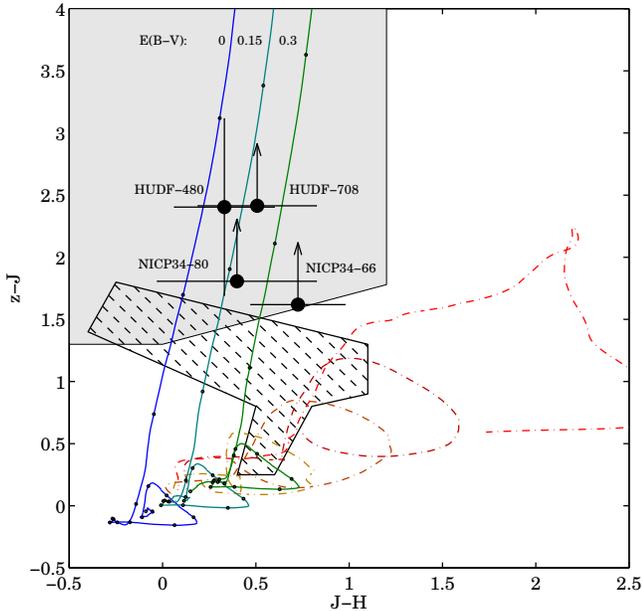}
	\caption{The $z-J$ vs.\ $J-H$ color-color diagram, on which we highlight the selection criterion for  $z>6.8$ galaxies (gray shaded area). Evolutionary tracks of star-forming galaxies, obtained with Bruzual \& Charlot (2003) models,  are shown with solid lines in blue-to-green colors, corresponding to dust obscurations of $E(B-V)=0,\ 0.15,\ 0.3$. Low redshift galaxy tracks,      up to redshift  $z=4$, are obtained from the galaxy templates of \citet{cww80}, and  are shown as dash-dotted tracks in orange-to-red colors. Dwarf stars are expected to lie  in the hatched region (Knapp et al. 2004, Burrows et al. 2006, Pickles 1998). The filled black circles correspond to the four $z\sim7$ candidate galaxies identified in our UDF05+HUDF dataset.}
	\label{fig:JH_ZJ}
\end{figure}

This selection is indicated as a gray shaded region in the $z-J$ vs.\  $J-H$ color-color diagram of  Fig. \ref{fig:JH_ZJ}. The figure also shows  tracks of different galaxy types as a function of redshift. A star-forming galaxy with metallicity $0.2Z_\odot$ and a stellar  age of $10^8$ yr is shown with three blue-to-green solid lines, corresponding to $E(B-V)=0,\ 0.15,\ 0.3$ mag, respectively (using the extinction law of Calzetti et al. 2000). Such star forming galaxies enter the selection window at $z\sim6.8$. Lower redshift galaxies are shown with dash-dotted lines in orange-to-red colors. The hatched region in the diagram corresponds to the expected location of cool dwarf stars of type M to L (Knapp et al. 2004\footnote{We have used the theoretical models of Burrows et al. 2006 to convert the photometry of Knapp et al.\  to the ACS/NICMOS filter set.}, Burrows et al. 2006, Pickles 1998).  

Note that the color selection that we have adopted in this study  is the so-called ``conservative'' criterion in \citet{bouw06}. Relative to the less restrictive color window that is used in B08, our selection criterion slightly reduces the effective selection volume by excluding  galaxies with redshifts between $z\sim6.5$ and  $z\sim6.8$; however, it is more robust against low redshift interlopers, and in particular it excludes more efficiently a possible contamination by passively-evolving  $z\sim1.7$ galaxies with a pronounced 4000 \AA\ break.

\begin{deluxetable*}{lccccccc}
\tablecaption{Photometry of the four \zdrop  sources\label{tab:phot}}
\tablewidth{0 pt}
\tablecolumns{8}
\tablehead{\colhead{ID} & $\alpha$ & $\delta$ &\colhead{\hF} &\colhead{S/N(\hF)} &\colhead{\jF-\hF} & \colhead{\zF-\jF} & 
 \colhead{FWHM}}

\startdata
HUDF-480  & 3:32:38.81 & -27:47:07.2 & $26.9\pm0.1$ & 7.2 &  $0.3\pm0.3$ & $2.4\pm0.7$ & 0\farcs30\\
HUDF-708  & 3:32:44.02 & -27:47:27.3 & $27.3\pm0.2$ & 6.8 &  $0.5\pm0.3$ & $>2.4 $     & 0\farcs29\\ 
NICP34-66 & 3:33:08.29 & -27:52:29.2 & $26.9\pm0.1$ & 12.5 & $0.7\pm0.3$ & $>1.6$      & 0\farcs54 \\ \hline
NICP34-80\tablenotemark{a} & 3:33:10.63 & -27:52:31.0 & $27.8\pm0.2$ & 4.8  & $0.4\pm0.4$ & $>1.8$  & 0\farcs35 
\enddata

\tablenotetext{a}{This object could be a spurious detection. We have  detected it with a 2$\sigma$ significance in  \jF, and it has not been confirmed by  an independent reduction of the same data by R. Bouwens (2008, private communication).}
\end{deluxetable*}

\begin{figure}[tbp]
	\centering
		\includegraphics[width=\linewidth]{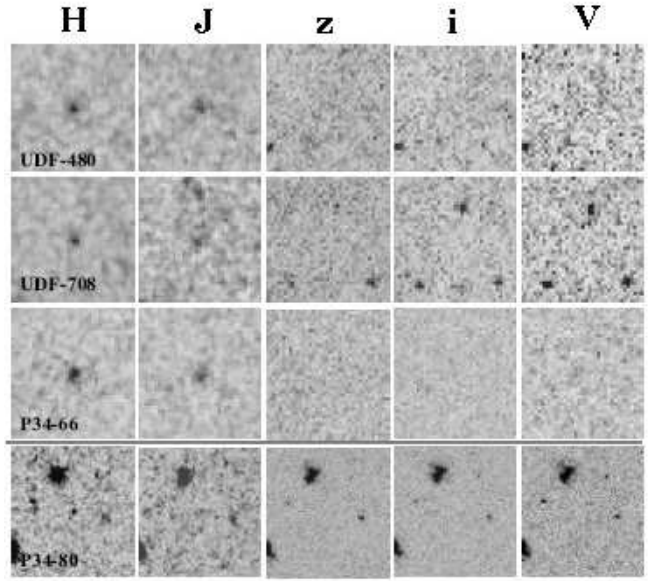}
	\caption{Images of our four \zdrop candidates. The boxes are 3\farcs5 in size. Left to right, the columns show images in  \hF, \jF, \zF, \iF, and \vF. The sources are very compact. All but the faintest candidate (NICP34-80) are securely detected both in \hF\ and \jF.}
	\label{fig:stamps}
\end{figure}

All SExtractor sources that satisfy the above color criterion were visually inspected; four $z>6.8$  LBG candidates were identified after removal of spurious detections (e.g., stellar diffraction spikes, edge artifacts, etc.). These four high-$z$ galaxy candidates  are shown as  filled black circles in the color-color diagram of Fig. \ref{fig:JH_ZJ}. Images of these sources are shown in Fig. \ref{fig:stamps}; their photometry is listed in Table \ref{tab:phot}. All four $z$-dropouts  are rather  faint ($H_{160}\gtrsim27$) and very compact (they are essentially unresolved at the resolution of the NICMOS images).  Two of the candidates lie in the shallow flanking region of the HUDF; the remaining  two are located in the shallower area of the NICP34 data. As evident from Fig. \ref{fig:JH_ZJ}, three out of our four high-$z$ candidates support relatively small amount of dust, $E(B-V)=0.15$,
while the last one (NICP34-66) is consistent with slightly larger extinction.

Three of our four sources are securely detected in both \jFilter and \hF. The faintest source (NICP34-80) has a S/N of 4.8 in the 0\farcs3 \hFilter aperture, and of only $\sim$2 in the \jFilter aperture. Thus NICP34-80 is a less secure high-$z$ candidate.  With the exclusion of NICP34-80, the remaining three  objects are also found in the independent analysis of B08. One of the $z_{850}$-dropouts in our analysis (HUDF-480) had already been reported in previous works \citep{bouw04b,bouw06,coe06}, and it has also been detected in Spitzer images \citep{labb07}, with infrared colors in very good agreement with a star-forming galaxy at $z\sim7$. The remaining $z_{850}$-dropouts in our sample are either undetected ($<2\sigma$; NICP34-66/80)) or severely blended (HUDF-708) in the existing, publicly available Spitzer data (from the IRAC followup of GOODS for the HUDF\footnote{http://data.spitzer.caltech.edu/popular/goods/} and from the SIMPLE survey for NICP34\footnote{http://data.spitzer.caltech.edu/popular/simple/}). The magnitude limits for the sources in NICP34 are quite bright and thus do not add any constraint on their photometric redshifts.

\subsection{Sources of sample contamination}

There are  five possible sources of contamination to the derived sample of $z\sim7$ galaxy candidates; these are due to  $(i)$ spurious detections, $(ii)$ cool dwarf stars, $(iii)$ low redshift galaxies with colors similar to $z\sim7$ galaxies, $(iv)$ high redshift supernovae, and $(v)$ photometric scatter of low redshift galaxies into our selection window. In detail:

$(i)$ Except for the faintest of our sources, NICP34-80, all candidates are detected at more than $6\ \sigma$ in \hFilter and $>3\sigma$ in \jF. It is therefore very unlikely that these high S/N sources are produced by peaks in the noise. In order to estimate the reality of NICP34-80, which is only detected at $2\ \sigma$ in \jF, we repeated our analysis on `negative images', obtained by multiplying the real images by $-1$ (after masking out all detected sources). One `negative source' is detected with  S/N(\hF)$>4.5$, S/N(\jF)$>2$ and a pixel flux distribution similar to NICP34-80. Thus, we conclude that NICP34-80 is likely a spurious detection.

$(ii)$ Except for NICP34-66, whose profile has a full-width-half-maximum (FWHM) of 0\farcs54, all sources are unresolved (FWHM $\sim 0\farcs29-0\farcs35$; see Table \ref{tab:phot}). For comparison, stars are measured to have FWHM of $0\farcs30-0\farcs39$ \citep[see also][]{thom05}.
Thus, on the basis of their profiles, we cannot exclude the possibility that our candidates are stars. However, they all show colors redder than expected for ultracool dwarfs. Such stars should have at least weak detections in our ultra-deep \zFilter data. Therefore it is unlikely that our candidates are dwarf stars. 
However, simulations of dwarf star LFs by \citet{burg04} predict that there will be $\sim 0.5$ stars of type L0-T8 in our survey in the magnitude range of $H_{AB}=27-28$. Thus, this is a non-negligible source of contamination.

$(iii)$ Galaxies at $z\sim1.2-1.8$ which emit  a combination of a red continuum (due to dust or old stellar age) and strong [OIII]$\lambda$5007,$\lambda$4959 and H$\beta$ emission lines can exhibit colors which place them within our selection window. The equivalent widths of the emission lines that are required are however rather extreme ($\gtrsim500$ \AA\ rest-frame), which makes this a rather implausible source of contamination.

$(iv)$ The 2.2 arcmin$^2$ NICMOS observations for the UDF05 data were taken at a much earlier epoch than the corresponding ACS images;  high redshift SNe could therefore also contaminate our sample. However, with a similar  calculation as in B08,  we expect to find only 0.03 SNe within the area of our NICMOS observations; this is  thus  a negligible source of contamination in our sample.

$(v)$ Photometric scatter could in principle  cause some contamination as well. In particular, $z\sim1.7$ galaxies with prominent 4000 \AA\ breaks, the closest to our selection window, could be a significant source of contamination. In order to assess the importance of this effect, we run simulations in which we applied a suitable photometric scatter to the HUDF source catalog of  \citet{coe06}. This catalog contains sources separately identified  both in the optical and in the infrared, and  includes one of our $z\sim7$ candidates. The catalog also  contains photometric redshifts for all sources. We can thus study the redshift distribution of galaxies which enter our selection window after applying the photometric scatter. 
Certainly this test is limited by the reliability of the photometric redshifts. However, it provides a worthy cross check on our candidates, and we found no low redshift source in any of the simulation runs. In general, requiring that an object is not detected in any of the optical bands efficiently reduces the contamination from lower redshift interloper galaxies.

In summary, three of the four detected $z_{850}$-dropouts are very likely to be genuine $z\sim7$ LBGs. 
However, we stress that future observations will be needed to confirm these candidates as high-z sources. 
In the following we discuss the implications for galaxy formation and reionization of assuming that these three objects are truly galaxies at $z\sim7$; we also separately comment on the impact of including the fourth, less likely high-$z$ candidate in our $z\sim7$ LBG sample.

\section{Results}
\label{sec:discussion}

\subsection{Expected vs. detected number of sources}
\label{sec:sim}

In order to estimate the completeness, $C$, and redshift selection probability, $S$, for our sample, 
we used the procedure that we outlined in \citet{oesch07}, i.e.,  we simulated data by inserting artificial galaxies into the real images, and run  SExtractor on the simulated data with the same parameters adopted to extract the original catalogs.  We adopted colors according to a Gaussian distribution of continuum slopes that is measured for $z\sim6$ LBGs ($\beta=-2.2 \pm 0.2$; Stanway et al. 2005), and a log-normal size distribution with a size scaling of $(1+z)^{-1}$ \citep{ferg04,bouw04}. 

Given a LF, the number of objects expected in a bin of a given magnitude, $\phi$,  is given by
\[
N(m)= \int_{\mathrm{bin}}dm' \int_0^\infty dz \frac{dV}{dz} \phi(M[m',z]) S(z,m') C(m')
\]
where $M[m,z]=m-K(z)-DM(z)$ is the absolute magnitude, $m$ the observed magnitude, $K(z)$ the $K$-correction term and $DM$ the distance modulus. We adopted a $10^8$ yr old, $0.2Z_\odot$, star-forming template for the $K$-correction from the observed  \hF-band to rest-frame $2000$ \AA, with $E(B-V)=0.15$, consistent with best-fit SEDs for $z\sim5$ LBGs \citep[e.g.][]{verm07}.  
Note that correcting to 1400 \AA\ results in a difference of only 0.07 mag. The total number of expected $z\sim7$ galaxies in our survey is then just the sum over all the fields, which all have different selection probabilities and completenesses due to the different depths and data quality.

Assuming no evolution in the LF from $z\sim6$ to 7, we would expect to detect 12 galaxies in our dataset (using the $z\sim6$ parameters of B07); in contrast, only three, or at most four (if the least reliable candidate, NICP34-80, is also included), are detected. In particular, we would expect about four galaxies brighter than $H_{160}<26.75$. No such bright candidate is observed, which leads us to conclude, albeit with a limited significance,  that  there is  evolution in  the LBG LF from $z\sim6$ to $z\sim7$.  From Poissonian statistics, the detection of four sources out of the expected 12 has a significance of 2.3 $\sigma$; when cosmic variance is considered\footnote{Our cosmic variance calculator is publicly available at http://www.stsci.edu/$\sim$trenti/CosmicVariance.html}, 
the significance of the result is reduced down to 1.4 $\sigma$. The same, i.e. a chance of 8\%, is found from accurate beam tracing through a dark matter simulation in which halos are populated with galaxies taking into account the specific geometry of our survey \citep{tre07}.
Note however  that, as commented above, the NICP34-80 source is likely a spurious detection; should this or any other high-$z$ candidate in our sample turn out to be an artifact (or a lower redshift interloper), this would imply a stronger evolution of the LF over the $z\sim6$ out to $z\sim7$ period. The estimates above therefore provide a lower limit to the real evolution of the LF across the reionization boundary.

\subsection{The $z\sim7$ LBG luminosity function}

The approach that we use to constrain the $z\sim7$ LBG LF from our data is to  fix the faint-end slope $\alpha$ and the normalization $\phi_*$, and search for evolution in  $M_*$  only; this approach is motivated by the finding that $M_*$ is the only LF parameter which substantially varies over the $z\sim4-6$ redshift range (B07). We adopt a linear evolution $M_*(z) = M_*(z=5.9)+\beta_*(1+z)$ and fit the only free parameter, $\beta_*$, to our four (three) $z\sim7$ candidates. We use the parameters of B07 to anchor the evolution at redshift 5.9, i.e.  we adopt $\phi_*(z=5.9)=1.4\times10^{-3}$ Mpc$^{-3}$mag$^{-1}$, $\alpha(z=5.9)=-1.74$, and $M_*(z=5.9)=-20.24\pm0.19$. 

To determine $\beta_*$, we maximize the likelihood
\[
\like= \prod_i {\cal P}[N_{\rm obs}(m_i),N_{\rm exp}^{\beta_*}]
\]
where ${\cal P}[x,\lambda]$ is the Poisson distribution with mean $\lambda$ evaluated at $x$; $N_{\rm obs}(m_i)$ is the observed number of objects in the magnitude bin $m_i$, and $N_{\rm exp}^{\beta_*}$ are the expected number of objects for a given evolution parameter $\beta_*$.

\begin{figure}[tb]
	\centering
		\includegraphics[width=\linewidth]{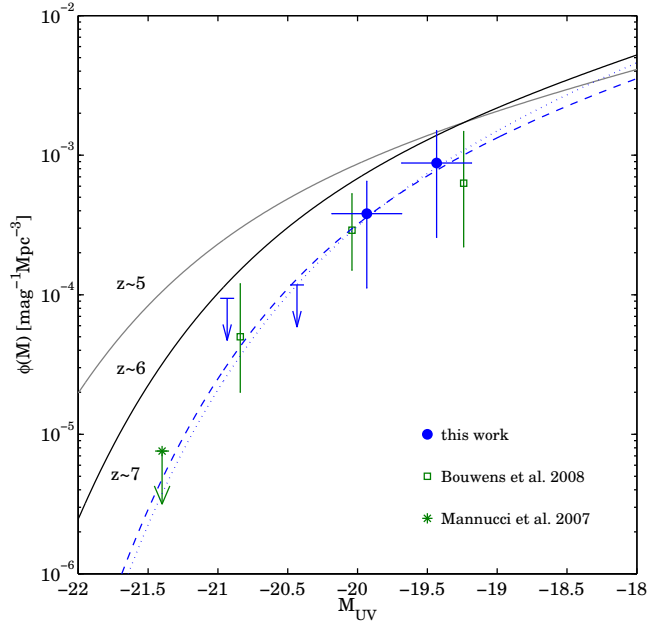}
	\caption{The $z\sim7.2$ LBG LF derived from our four candidates (filled blue circles). The best-fit Schechter function, derived from maximizing the likelihood for the evolution parameter $\beta_*$, is shown as a dashed line. The dotted line is the same, but with a steeper faint end slope $\alpha=-1.9$. Upper limits for empty magnitude are indicated with blue arrows. Open squares are from Bouwens et al. (2008), with whom we have three high-$z$ candidates in common. The upper limit at $M=-21.4$ is from \citet{mann06}. The other lines show the LF at $z\sim5$ (gray solid line; Oesch et al. 2007), and $z\sim6$ (black solid line; Bouwens et al. 2007).}
	\label{fig:LFevol7}
\end{figure}

The best fit  for $\beta_*$ is $0.36\pm0.18$, which translates into $M_*(z=7.2)=-19.77\pm0.30$. If we exclude the faintest, least reliable \zdrop candidate from the fit, the result changes only slightly, i.e., $\beta_* = 0.43\pm0.19$, or $M_*(z=7.2)=-19.68\pm0.31$.
Within the errors these values are identical, and we adopt a fiducial value of $M_*(z=7.2) = -19.7\pm0.3$ at 2000 \AA. 
With this evolving LF, the redshift distribution of our sources is predicted to be $z=6.8 - 7.7$ (within the 16th and 84th percentiles), with a mean $\overline{z}=7.2$. The resulting LF is shown in Fig. \ref{fig:LFevol7} as a dashed line; the figure  assumes that all candidates lie at $z\sim7.2$, with the effective volume estimated as $V_{\rm eff}(m) = \int_0^\infty dz \frac{dV}{dz} S(z,m)C(m)$.

Since the measured evolution is mostly driven by the non-detection of bright sources, the effect of varying the faint-end slope $\alpha$ is very small. Varying $\alpha$ by 0.16 (the formal 1-$\sigma$ error of the $z\sim6$ LF)  changes the value of  $M_*(z=7.2)$ by only $\pm 0.03$ mag. We have also tested how the LF would evolve if  $M_*$ does not evolve while $\phi_*$ varies with redshift; this would result in a $\phi_*$ value a factor $\sim 2.5$ lower than at $z\sim6$, i.e., $\phi_*=(0.56\pm0.50) \times10^{-3}$.

B08 used a large sample of NIR data, covering an area of 261 arcmin$^2$ (including the ultra-deep NICMOS data of this work) and found $M_*=-19.8$ and  $\phi_*=1.1\times10^{-3}$ mag$^{-1}$Mpc$^{-1}$, in very good agreement with our result (see also Fig. \ref{fig:LFevol7}). Furthermore, \citet{mann06} searched the ground-based VLT/ISAAC $J$, $H$, and $Ks$ data on the GOODS field for $z_{850}$ dropouts and, after removing probable dwarf stars, they found no candidates down to a UV rest-frame absolute magnitude of  $M_{1500} = -21.4$. These authors inferred an upper limit on the LF down to this magnitude which we also report  in Fig. \ref{fig:LFevol7}; this upper limit is in good agreement with the extrapolation to brighter magnitudes of our $z\sim7$ LF.

\section{Implications for Galaxy Evolution and Reionization}
\label{sec:discussion2}

\subsection{Evolution of the star formation rate density and  star formation histories of $z\sim6$ LBGs}

From the above LF we can compute the amount of evolution in the cosmic star formation rate density from $z\sim6$ out to $z\sim7$. Integrating down to $0.2L_*(z=6)$, corresponding to $M=-18.5$, we find a decrease in the luminosity density by 50\%, i.e., $\log(\rho[\mathrm{erg/s/Hz/Mpc}^3])=25.5\pm0.2$. Using the relationship between UV continuum and star formation rate of  \citet{kenn98}, this corresponds to a star formation rate density of $10^{-2.32\pm0.23}$ M$_\odot$ yr$^{-1}$Mpc$^{-3}$.
Assuming that the LF evolves only in density implies instead a reduction of 60\%. Thus, independent from the specific assumption on the LF evolution (either in $M_*$ or in $\phi_*$), down to this brightness limit the luminosity density, and thus the star formation rate density, drops by a factor of $\sim2 - 2.5$ over the $\sim$ 170 Myr that separate the $z=6$ and $z=7$ epochs. \citet{mann06} and B08 independently  find similar results. 
Note that the conversion of Kennicutt is applicable for galaxies with continuous star-formation over timescales of a few 100 Myrs and Salpeter initial mass function (IMF). They might differ for galaxies which are experiencing stochastic bursts, as is assumed for high-$z$ LBGs \citep{verm07}. However, larger differences are expected due to uncertainties in the IMF. 

The observed decline of the bright end of the LF from $z\sim6$ out to $z\sim7$ provides an upper boundary on the ensemble average star formation rate of  $z=6$ LBGs  $\sim170$ Myr  prior the epoch at which  they are observed.  Eyles et al.\  (2007) have  investigated the ensemble average star-formation history of $i_{775}$-dropouts brighter than $z_{850}=27$ in the GOODS field, and  suggested that this rises from $z\sim6$ to a peak at $z\sim 7.5$, and declines beyond this redshift.  Using  Bruzual \& Charlot (2003) models and assuming the  star formation history of \citet{eyle07}, we find that the rest-frame 1500 \AA\ luminosity density should show an increase by a factor of $\sim 2.5$ from $z\sim6$ up to 7, i.e., in striking contrast with the  drop of the luminosity density that we report above. 
Dust corrections can not reconcile this disagreement, unless an evolution towards larger dust obscuration at higher redshifts is invoked. 

This apparent contradiction could however be an artifact of the specific star formation history that has been adopted by Eyles at al.\ in the computation of the ensemble average. Generally, Spitzer data of  $i_{775}$-dropouts in the GOODS fields  have revealed prominent $4000$ \AA\ breaks in a large fraction of these sources  ($\sim 40\%$), indicating up to 90\% of old stellar populations in those $z\sim6$ galaxies \citep{yan05,eyle07}.  Our results on the $z\sim7$ LF  provide global support to a scenario where a large fraction of stars in the $z\sim6$ population form at redshifts well above $z\sim7$.  

\subsection{Do galaxies reionize our universe?}

Consistent with the decrease in cosmic star formation rate density, the rate of ionizing photons produced by LBGs must also decrease between $z\sim6$  and $\sim7$.

\begin{figure}
	\centering
		\includegraphics[width=\linewidth]{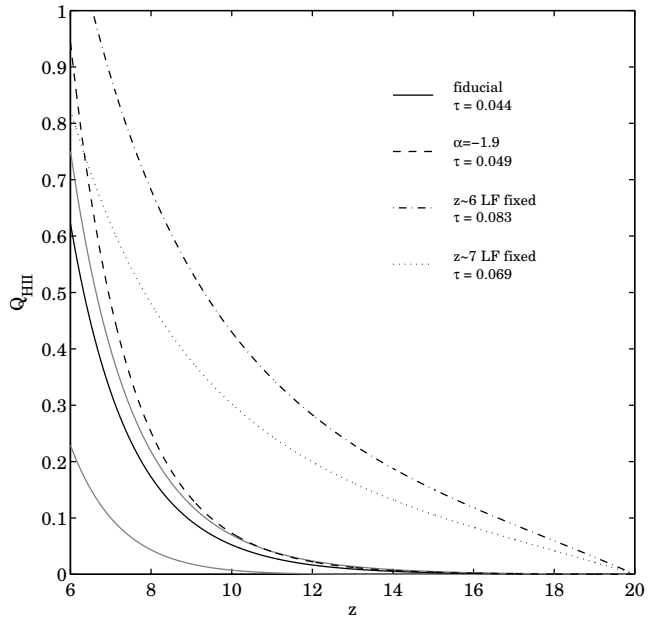}
	\caption{Filling factor of ionized hydrogen as a function of redshift. The solid lines show the evolution of the filling factor calculated by assuming a faint end slope $\alpha=-1.74$ and an evolution of the galaxy LF by 0.35 mag per unit redshift, and by integrating down to different limiting SFRs ($\sim1,\ 0.01,\ 0.0001$ M$_\odot/$yr; lower gray, black, and upper gray lines, respectively). The dashed line assumes the same evolution in $M_*$ but $\alpha=-1.9$. The dash-dotted line corresponds to no evolution of the LF from $z\sim6$ out to $z=20$; the dotted one assumes the same but with the $z\sim7$ LF. If not stated otherwise all curves are integrated down to SFRs of $10^{-2}$ M$_\odot/$yr. The corresponding electron scattering optical depths are also indicated.}
	\label{fig:QH2_evolvingLF}
\end{figure}

\citet{bolt07} pointed out that, with the estimated $z\sim7$ LF of \citet{bouw04b}, the number of ionizing photons falls short by almost an order of magnitude even to just keep the universe ionized at $z\sim7$. Only models in which the star formation rate rises or stays constant beyond $z\sim6$ are able to provide enough ionizing photons. This result was however derived by including only  galaxies with absolute magnitudes brighter than $M=-18$ mag; this corresponds to a lower limit for the star formation rate of $\sim 1$ M$_\odot/$yr (see lower gray line in Fig. \ref{fig:QH2_evolvingLF}). For LF faint-end slopes as steep as those that are measured, however, $\sim 50\%$ of the luminosity density  comes from galaxies beyond the current detection limits at $z\sim6$ ($M>-17$ mag). Thus, photons from these very faint  galaxies will add a substantial contribution to the reionizing flux. 

Here we adopt the results of our independent analysis of the $z\sim7$ LF, which, using substantially improved data products,   confirms the drop in the star formation rate density from $z=6$ to $z=7$ suggested by \citet{bouw04b}. We follow the approach of \citet{bolt07} to estimate the contribution of galaxies to reionization.  We include, however, the contribution of  faint galaxies to reionization by decreasing the lower limit star formation rate to $10^{-2}$ M$_\odot/$yr in our calculations. 

Computations of the filling factor of ionized hydrogen, $Q_\mathrm{HII}$, critically depend on the adopted values of the clumping factor $C$ and escape fraction $f_\mathrm{esc}$. Scenarios with larger clumping factors lead to more efficient recombinations and thus require more ionizing photons;  lower escape fractions also require more photons. Both the clumping factor and the escape fraction  are however  poorly known. To give an upper limit on the potential for reionization of the galaxy population, we adopt (possibly somewhat optimistic) values of $C=2$, as found in simulations \citep[see discussion in][]{bolt07},  and $f_\mathrm{esc}=20\%$, the  upper limit that has been inferred from both  observations and simulations \citep[e.g.][]{shap06,sian07,gned08a}. We neglect any possible mass dependence of $f_\mathrm{esc}$.

Assuming $(i)$ that the constant dimming of $M_*$ with redshift continues beyond $z\gtrsim7$ and $(ii)$ a faint-end slope $\alpha$ which remains constant at the value  $\alpha=-1.74$ that is measured at $z\sim6$ (i.e., our ``fiducial" evolution parameters),  the galaxy population would be able to ionize the universe only up to a fraction $Q_\mathrm{HII}\sim 65\%$ (see solid lines in Fig. \ref{fig:QH2_evolvingLF}). The corresponding optical depth to electron scattering  is $\tau = 0.044$, which is $\sim2.5\sigma$ lower than the WMAP-5 value of $\tau = 0.087\pm0.017$ \citep{dunk08}. Thus, this reionization scenario starts too late, ionizing too few atoms at high redshifts. Indeed the WMAP-5 data  suggest that the Universe underwent an extended period of partial reionization, beginning as early as $z\sim20$, similar to the double reionization scenario proposed by \citet{cen03}. If by $z\sim9$ the Universe is already substantially ionized by some other sources, the galaxy population would certainly be able to complete the reionization process by $z\sim6$. 

Another possibility is that the $z>7$ LF has a faint-end slope which is steeper than currently observed at $z\sim6$ \citep[predicted e.g. by][]{ryan07}. The current estimate(s) do not allow one to rule out that the faint end of the LF at such early epochs is as steep as $\alpha=-2$, which would result in a diverging luminosity density. To assess whether the faint galaxy population can provide enough additional photons to complete reionization by $z\sim6$, we adopt a value of $\alpha=-1.9$ (see dotted line in Fig. \ref{fig:LFevol7}) and recompute the filling factor of neutral hydrogen under these modified assumptions. This scenario is shown in Fig. \ref{fig:QH2_evolvingLF} as a dashed line. Such a steep faint-end slope for the LF  would result in an end of reionization by $z\sim6$, and in an integrated electron scattering optical depth of $\tau=0.049$, which is still substantially lower than the current WMAP-5 value.  For comparison, we also show in Fig  \ref{fig:QH2_evolvingLF} the evolution of the HII filling factor assuming no evolution in the LF from $z\sim6$ ($z\sim7$) to $z=20$. Integrating the luminosity density again to $10^{-2}$ M$_\odot/$yr, this results in the completion of reionization by $z\sim6.6$ ($\sim5.4$) and in an optical depth $\tau=0.083$ ($0.069$).

We therefore conclude that: $(i)$ In order for the galaxy population to provide a substantial contribution to reionization, galaxies below current detection limits must play a significant role; $(ii)$ Under this assumption, a galaxy population which evolves by 0.35 mag per redshift unit would be able to ionize the universe by $z\sim6$, provided that the LF faint-end slope is steeper than the $\alpha=-1.74$ value that is measured at redshifts $z\lesssim6$; $(iii)$ Even under this assumption, the resulting optical depth to electron scattering is lower by $\sim2 \sigma$ relative to the WMAP-5 measurement, indicating that too few atoms are ionized at very high redshift. This finding provides evidence that, if reionization is primarily driven by galaxies, either the currently detected evolution of the galaxy population slows down at $z\gtrsim7$, or the LF evolution is compensated by a decrease in metallicity and a corresponding increase in ionization efficiency \citep{stia04}. Otherwise, other sources must substantially participate to reionize our Universe like miniquasars or population III stars \citep[e.g.][]{shul08,mada04,soka04}.

\acknowledgments
Acknowledgements: We wish to thank the anonymous referee for helpful comments that have improved the presentation of these results. PO thanks Rychard Bouwens for helpful discussions and for private communication of the result of an independent reduction of the data to establish the reality of our faintest candidate. This work has been partially supported by NASA HST grant 01168.

Facilities: \facility{HST(ACS/NICMOS)}.

\end{document}